\documentclass[aps,pre,twocolumn,groupedaddress]{revtex4}
\usepackage{graphicx}
\usepackage{amsmath}

\begin{document}

\title{Duality analysis on random planar lattice}
\author{Masayuki Ohzeki$^1$ and Keisuke Fujii$^2$}
\date{\today}

\address{$^1$Department of Systems Science,
Graduate School of Informatics, Kyoto University, Yoshida-Honmachi,
Sakyo-ku, Kyoto, 606-8501, Japan}

\address{$^2$Graduate School of Engineering Science, Osaka University, Toyonaka, Osaka, 560-8531, Japan}

\begin{abstract}
The conventional duality analysis is employed to identify a location of a critical point on a uniform lattice without any disorder in its structure.
In the present study, we deal with the random planar lattice, which consists of the randomized structure based on the square lattice.
We introduce the uniformly random modification by the bond dilution and contraction on a part of the unit square.
The random planar lattice includes the triangular and hexagonal lattices in extreme cases of a parameter to control the structure.
The duality analysis in a modern fashion with real-space renormalization is found to be available for estimating the location of the critical points with wide range of the randomness parameter.
As a simple testbed, we demonstrate that our method indeed gives several critical points for the cases of the Ising and Potts models, and the bond-percolation thresholds on the random planar lattice.
Our method leads to not only such an extension of the duality analyses on the classical statistical mechanics but also a fascinating result associated with optimal error thresholds for a class of quantum error correction code, the surface code  on the random planar lattice, which known as a skillful technique to protect the quantum state.
\end{abstract}

\maketitle
\section{Introduction}
The duality, which bridges two distinct concepts or situations, is often found and plays important roles in various realms of physics and mathematics.
We focus on the duality in statistical mechanics, which is embedded in the symmetry of the partition function at low and high temperatures.
This symmetry allows us to identify the locations of the critical points for various spin models such as the Ising and Potts models \cite{Kramers1941,Wu1976}.

The most remarkable development on the duality in the spin systems is the successful application to the random spin systems, such as the spin glass, which has phase diagrams of complicated structures and exhibits rich behaviors. 
Since the series of work \cite{Nishimori2002,Mailard2003,Takeda2005,Nishimori2006} revealed that the duality is capable to estimate the approximate values of the critical points even for such complicated systems, the duality has been developed as one of the rare successful approaches to finite-dimensional spin glasses, where very little analytical work exists.
More precise analysis of the locations of the critical points in spin glasses has been performed by employing a real-space renormalization technique, i.e., the partial summation of the degree of freedom \cite{Ohzeki2008,Ohzeki2009a}.

Very recently, the duality also plays an important role to provide a powerful tool in other fields. 
For instance, it has opened a way to analyze the theoretical limitation in the classical/quantum coding theory, 
where order-disorder phase transition corresponds to the theoretical limitation of the error correction codes. 
Specifically, the optimal error thresholds of a class of quantum error correction codes, so-called surface codes \cite{Dennis2002}, have been evaluated in terms of the duality with the real-space renormalization \cite{Ohzeki2009c,Hector2012},
On the other hand, newly developed quantum error correction codes \cite{Loss2012} also motivate us to study the associated spin glass models, which have not been considered so far.

In the present paper, we extend the conventional duality analysis to a broader class of the lattices, which have various types of randomness.
Specifically, we deal with random planar lattices generated by random modification, bond dilution and contraction.
The random planar lattice consists of a mixture of the triangular and the hexagonal lattices through a single parameter to control its structure.
Since the conventional duality analysis is limited to a few self-dual lattices (e.g. the square lattice), we cannot deal with such a random planar lattice on all parameter values systematically.
In order to overcome this situation, we establish a unified and systematic approach to analyze the random planar lattices
by use of the duality with the real-space renormalization.
In certain extreme cases, we can reproduce the critical points on the triangular and hexagonal lattices, where the so-called star-triangle transformation works well with the duality analysis \cite{Domb1972,Wu1982}.
In this sense, the present framework can be regarded as a natural extension of the star-triangle transformation to a broader range of the random planar lattices.

\section{Duality analysis}
The structure of the random planar lattice is based on the square lattice, where we can obtain the exact location of the critical points via the duality analysis by virtue of its special symmetry called self-duality \cite{Kramers1941,Wu1976}.
Let us here briefly review the duality analysis by taking a simple case.
We consider the $q$-state Potts spin system with the following Hamiltonian
\begin{equation}
H = -\sum_{\langle ij \rangle}J_{ij} \delta\left(S_i,S_j\right),
\end{equation}
where $\delta(x,y)$ is Kronecker's delta and $S_i$  takes $0,1,\cdots,q-1$.
Now, we set the couplings $J_{ij}$ to be a positive value $J>0$ for simplicity.  
The duality is a symmetry argument in the low and high-temperature expansions of the partition function $Z=\sum_{S_i}\prod_{\langle ij \rangle}\exp(\beta J \delta(S_i,S_j))$ \cite{Kramers1941}.
From a modern point of view, these expansions can be related with each other by the $q$-component discrete Fourier transformation of the local part of the Boltzmann factor, namely edge Boltzmann factor $x_k = \exp(\beta J \delta(k,0))$ for a single bond \cite{Wu1976}.
Each term in the low-temperature expansion can be expressed by $x_k$.
On the other hand, the high-temperature expansion is expressed by the dual edge Boltzmann factor through the discrete Fourier transformation $x_l^*=\sum_{k} x_k \exp(i2\pi k l/q)/\sqrt{q}$.
By using this fact we can obtain a double expression of the partition function as
\begin{equation}
(x_0)^{N_B}z(u_1,u_2,\cdots) = (x^*_0)^{N_B}z(u^*_1,u^*_2,\cdots),\label{Zeq}
\end{equation}
where $z$ is the normalized partition function $z(u_1,u_2,\cdots)=Z/(x_0)^{N_B}$ and $z(u^*_1,u^*_2,\cdots)=Z/(x^*_0)^{N_B}$.
We here define the relative Boltzmann factors $u_k = x_k/x_0=\exp(-\beta J)$ and $u_k^*= x^*_k/x^*_0=(\exp(\beta J)-1)/(\exp(\beta J)+q-1)$.
The well-known duality relation $\exp(\beta^* J)=(\exp(\beta J)+q-1)/(\exp(\beta J)-1)$ is obtained by rewriting $u^*_k$ by $u_k$ with the different coupling constant, which implies the transformation of the temperature.
Its fixed point condition $\exp(\beta_c J)=(\exp(\beta_c J)+q-1)/(\exp(\beta_c J)-1)$ gives the critical temperature.
We can evaluate the exact value of the internal energy and the rigorous bound for the specific heat at the critical point from the relation of the partition function (\ref{Zeq}) through the appropriate derivatives as well as its location \cite{Nishimori2011}.

If we are interested in derivation of the location of the critical point, we can use an alternative way. 
We impose that both of the principal Boltzmann factors $x_0$ and $x_0^*$ with edge spins parallel are equal, (i.e. $x_0=x_0^*$).
We reproduce the equation $\exp(\beta_c J)=(\exp(\beta_c J)+q-1)/(\exp(\beta_c J)-1)$.
Notice that the symmetry of the low- and high-temperature expansions on the partition functions is embedded compactly in the edge Boltzmann factors only on the single bond.
In particular, the above argument is extremely simplified if we use the principal Boltzmann factors.

In the duality analysis, two expansions are related in-between the original and its dual lattices. 
In the dual lattice, the site and plaquette of the original lattice are exchanged with each other, and the bonds are perpendicular to those on the original lattice.
The bonds described on the dual lattice are perpendicular to the original bonds. 
The square lattice is one of the self-dual lattices, which share the same structure even after the transformation.
Usually, the duality analyses are limited to the case on the self-dual lattice.
For non-self-dual lattices, such as the triangular and hexagonal lattices, the duality analysis gives only the relationship between critical points for mutually dual pair \cite{Wu1976,Wu1982,Nishimori2011}.

In the case we will consider, the coupling constants $J_{ij}$ are multivariated due to the existence of the quenched randomness on the interaction and/or structure of the lattice, which we will introduce later. 
This hampers the straightforward application of the conventional duality analysis.
We then employ the replica method, which generalizes the duality analysis to deal with the quenched randomness \cite{Nishimori2002,Mailard2003}.

Let us consider the duality for the replicated partition function as $[Z^n]$ (written simply as $Z_n$ below), where $[\cdots]$ is the configurational average for the quenched randomness according to the distribution functions of $J_{ij}$.
The $q^n$-component discrete Fourier transformation again leads us to the double expressions of the replicated partition function as
\begin{equation}
Z_n(x_0,x_1,\cdots) = Z_n(x^{*}_0,x^{*}_1,\cdots) \label{duality0}.
\end{equation}
We extract the principal Boltzmann factors from both side of Eq. (\ref{duality0}), and obtain
\begin{eqnarray}
(x_0)^{N_B}z_n(u_1,u_2,\cdots) = (x^*_0)^{N_B}z_n(u_1,u_2,\cdots),
\end{eqnarray}
where $z_n(u_1,u_2,\cdots)=Z_n(x_0)^{N_B}$ and $z_n(u*_1,u^*_2,\cdots)=Z_n]/(x^*_0)^{N_B}$.
In the case with the quenched randomness, unfortunately we cannot replace $u^*_k$ by $u_k$ as the previous case without the quenched randomness, since the replicated partition function is multivariable in general.
Nevertheless, we assume that $x_0$ is equal to $x^*_0$ again as done for the case without the quenched randomness, and take the limit of $n \to 0$ following the recipe of the replica method.
As a consequence, we estimate the approximate location of the critical points for spin glasses.

\begin{figure}[tbp]
\begin{center}
\includegraphics[width=60mm]{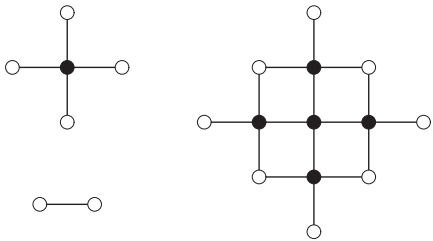}
\end{center}
\caption{The unit cluster for the square lattice with four bonds $s=1$ (left upper panel).
The left lower panel describes the single bond once we considered for the simple application of the conventional duality with the replica method.
The right panel denotes the larger cluster with $16$ bonds $s=2$ to identify more precise location of the critical points.
The black circles denote the spin variables summed over.
The white circles express the spin variables without summation.}
\label{fig1}
\end{figure}

By taking a wider range of the local part of the Boltzmann factor rather than a single bond, we can improve the estimates, and the resultant values are expected to asymptotically reach the exact solutions \cite{Ohzeki2008,Ohzeki2009a}.
For instance, in the case on the square lattice, we define the cluster Boltzmann factor $x_k^{(s)}$, where the subscript $k$ denotes the configuration of the edge (white-colored) spins, and $s$ distinguishes the size of the cluster as in Fig. \ref{fig1}.
Then we can rewrite the equality from the conventional duality (\ref{duality0}) as
\begin{equation}
Z_n(x^{(1)}_0,x^{(1)}_1,\cdots) = Z(x^{*(1)}_0,x^{*(1)}_1,\cdots) \label{duality1}.
\end{equation}
Again we extract the renormalized-principal Boltzmann factors $x^{(s)}_0$ and $x_0^{*(s)}$ from both sides.
\begin{eqnarray}\nonumber
&&(x^{(s)}_0)^{N^{(s)}_B}z^{(s)}(u^{(s)}_1,u^{(s)}_2,\cdots) \\
&& \quad = (x^{*(s)}_0)^{N^{(s)}_B}z(u^{*(s)}_1,u^{*(s)}_2,\cdots),
\end{eqnarray}
where $z^{(s)}$ is the normalized partition function defined as $z^{(s)}(u^{(s)}_1,u^{(s)}_2,\cdots)=Z_n/(x^{(s)}_0)^{N^{(s)}_B}$, and $z(u^{*(s)}_1,u^{*(s)}_2,\cdots)=Z_n/(x^{*(s)}_0)^{N^{(s)}_B}$.
We here define the renormalized-relative Boltzmann factors $u^{(s)}_k = x^{(s)}_k/x^{(s)}_0$ and $u_k^{*(s)}= x^{*(s)}_k/x^{*(s)}_0$.
We set the equation to estimate the more accurate location of the critical point than $x_0=x_0^*$ as inspired by the previous case without the quenched randomness,
\begin{equation}
x_0^{(s)} = x_0^{*(s)}, \label{MCP}
\end{equation}
where $x_0^{(s)}$ and $x_0^{*(s)}$ express the renormalized Boltzmann factor with the edge spins parallel as the detailed calculation shown below.
The conventional duality with the single bond can be understood as the special case $s=0$.

Although the above method is not exact, it is believed that the estimations systematically approach the exact solutions for the critical points of spin glasses by increasing the size of the cluster \cite{Ohzeki2008,Ohzeki2009a,Ohzeki2011a}.

\section{Random planar lattice}
\begin{figure}[tbp]
\begin{center}
\includegraphics[width=80mm]{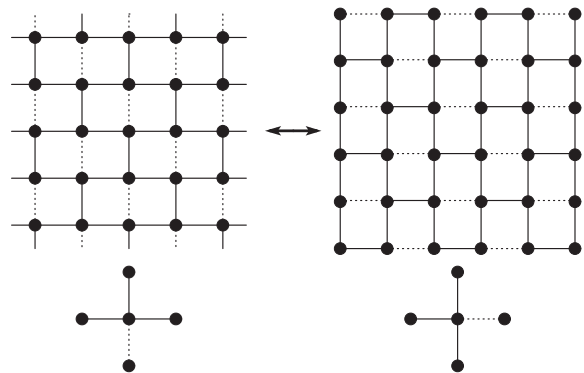}
\end{center}
\caption{Structure of the random planar lattice. 
The lower panels depict a part of the unit cell of the whole lattice.
The two types of modification, "bond dilution" and "bond contraction" are assigned on each dashed bond.
The right panels give the dual lattice, while the left ones express the original lattice.
Readers can confirm the self-duality holds on the structure of the lattice even with existence of a part of the modifications.
}
\label{fig2}
\end{figure}

\begin{figure}[tbp]
\begin{center}
\includegraphics[width=80mm]{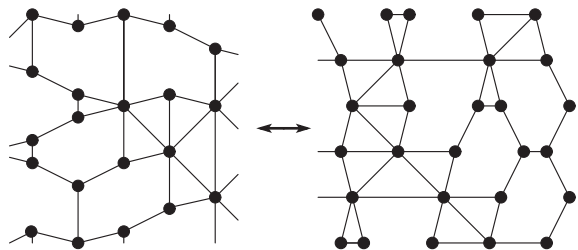}
\end{center}
\caption{One of the realizations of random planar lattice (left), and its dual one (right).}
\label{fig3}
\end{figure}

\begin{figure}[tbp]
\begin{center}
\includegraphics[width=80mm]{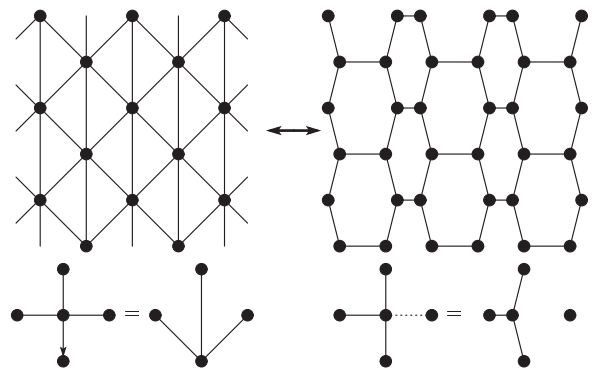}
\end{center}
\caption{Extreme limit of $p_{\rm mix.} = 0$.
The original lattice in the left panel becomes the triangular lattice.
On the other hand, the dual one in the right-hand side is the hexagonal lattice.
For guide to eyes, we rearrange the location of the sites as the modifications of the unit cell on the lower panels.
}
\label{fig4}
\end{figure}

\begin{figure}[tbp]
\begin{center}
\includegraphics[width=80mm]{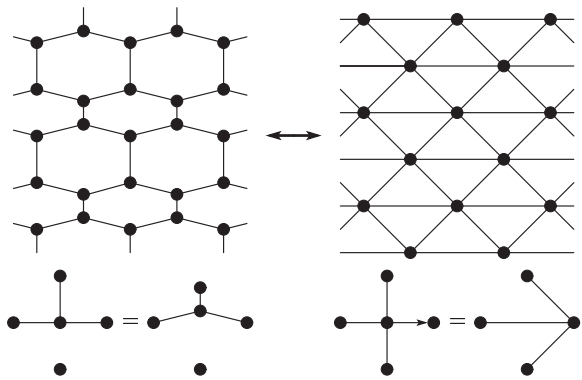}
\end{center}
\caption{Extreme limit of $p_{\rm mix.} = 1$.
The original lattice in the left panels turns into the hexagonal lattice, while the dual one in the right-hand side becomes the triangular lattice.
The same symbols are used as those in Fig. \ref{fig4}.
}
\label{fig5}
\end{figure}

Let us take a step toward the main issue in the present study.
We introduce two types of quenched randomness for the coupling constant $J_{ij}$ to form the random planar lattice.
One is the bond dilution, which means that we remove the bond.
The other is bond contraction by which we put two of the sites on the end of the bond together.
The bond dilution and contraction expressed by $J_{ij}=0$ and $J_{ij}=\infty$ are given with the probability $p_{\rm mix}$ and $1-p_{\rm mix}$, respectively.
We chose a single bond on each unit cell with such randomness as in Fig. \ref{fig2}.
We describe the representative instance of the random planar lattice with $p_{\rm mix}=1/2$ in Fig. \ref{fig3}.
If we take the extreme limit $p_{\rm mix} \to 0$, the resultant lattice becomes the triangular lattice as in Fig. \ref{fig4} (left).
On the other hand, when $p_{\rm mix} \to 1$, it becomes the hexagonal lattice as in Fig. \ref{fig5} (left).
These two lattices are known as a mutually dual pair.
While, in this case, we can obtain a relation between two critical points on the mutually dual pair of lattices, we cannot identify the location of the critical point.
Thus we needed to employ an additional tool to derive the location of the critical point, such as the star-triangle transformation \cite{Domb1972,Wu1982}.
However, once we regard the random planar lattice as the square lattice with quenched randomness such as $J_{ij} = 0$ and $J_{ij}=\infty$ given by probability $p_{\rm mix}$ and $1-p_{\rm  mix}$, respectively, we may perform the duality analysis on such a random spin model by virtue of the duality of the original square lattice.
Actually it will be shown below that the self-duality on the original square lattice always holds for any $p_{\rm mix}$.
This is the reason why we can employ the duality analysis with the real-space renormalization on the cluster of the unit cell as in Fig. \ref{fig1} to analyze the location of the critical point on the random planar lattice.

\begin{figure}[tbp]
\begin{center}
\includegraphics[width=80mm]{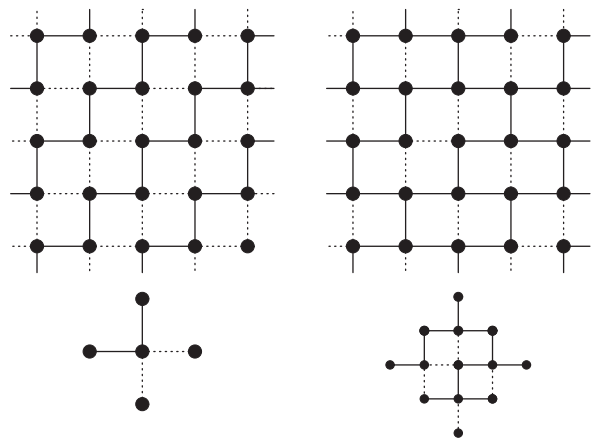}
\end{center}
\caption{Other random planar lattices where one can perform the duality analysis.
The left panel has a simple case by tiling of the unit cell with four bonds.
The right panel is a more complicated case by the unit cell with $16$ bonds.}
\label{fig6}
\end{figure}

Beyond the present case, we can generalize the conditions on the class of random planar lattices, where the duality works well by considering modification starting from the square lattice.
The key point is to hold the self-duality of the square lattice even for the location of the quenched randomness, i.e., bond dilution and contraction.
As in Fig. \ref{fig2}, the dual lattice has the same structure including the locations of the bonds to be modified as that of the original one.
One can confirm this fact by rotating the dual lattice by $90$ degrees. 
Following this rule, we demonstrate other random planar lattices by tiling the unit cells as in Fig. \ref{fig6}.
The first instance as in Fig. \ref{fig6} (left) is trivial as a spin model, since it does not undergo any phase transition in finite temperature.
The other as in Fig. \ref{fig6} (right) possesses a complicated structure, on which the spin models would show finite-temperature phase transitions.

\begin{figure}[tbp]
\begin{center}
\includegraphics[width=80mm]{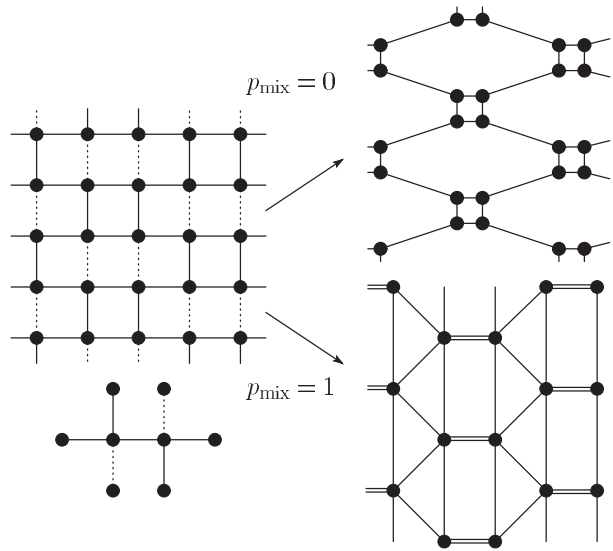}
\end{center}
\caption{A random planar lattice where one can not obtain the critical points.
In the lower left panel, we describe the unit cell.
The extreme limit $p_{\rm mix}=0$ gives the $(4,8^2)$ lattices.
This is the notation of the uniform tilings, in which all polygons are regular and each vertex is surrounded by the same sequence of polygons. 
The notation $(4,8^2)$ expresses that every vertex is surrounded by one square and two octagons.}
\label{fig7}
\end{figure}

\begin{figure}[tbp]
\begin{center}
\includegraphics[width=80mm]{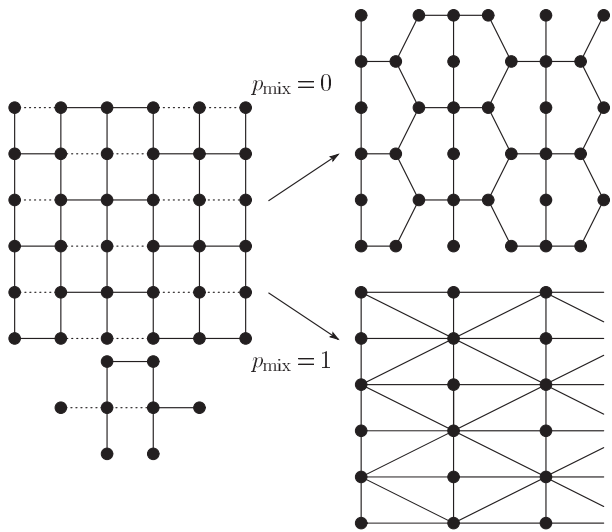}
\end{center}
\caption{The dual lattice of the random planar lattice depicted in Fig. \ref{fig7}.
The same symbols are used as those in Fig. \ref{fig7}.
The extreme limit $p_{\rm mix}=1$ becomes the Union-Jack lattices, which is the mutually dual pair of the $(4,8^2)$ lattice.
}
\label{fig8}
\end{figure}

One might imagine the generalization of the random planar lattice enables us to investigate most of the planar lattices by considering appropriate modifications.
However, in most cases, one cannot find such appropriate modifications.
The system on the random planar lattice possesses  inhomogenous interactions, if we regard the bond under dilution or contraction as the different interaction denoted by $J_{ij}=0$ or $\infty$ from the uniform ones $J_{ij}=J$.
Let us take a counterexample here.
We consider a random planar lattice by use of the unit cell as shown in Fig. \ref{fig7}.
The extreme limit $p_{\rm mix} = 0$ becomes the $(4,8^2)$ lattice.
In contrast to the successful instance, which relates the triangular and hexagonal lattices as shown in Fig. \ref{fig2}, the dual lattice described in Fig. \ref{fig8} is not the same structure of the original square lattice, since the location of the bond dilution/contraction in the unit cell differs from the original one.
Actually the extreme limit $p_{\rm mix} = 1$ for the dual random planar lattice is the same as the Union-Jack lattice, which is known as the mutually dual pair of the $(4,8^2)$ lattice.
Our duality analysis can give the connection between the critical points on the random planar lattices with $p_{\rm mix}$ and $1-p_{\rm mix}$ not only for the extreme case.
In this sense, we generalize the concept of the mutually dual pair into the case on the random planar lattice.
However we cannot estimate the accurate locations of the critical points on this type of the random planar lattice only by the duality analysis, since the random planar lattice does not hold self-duality on the original square lattice as seen above.
The duality analysis can be performed only on the random planar lattice that satisfies the self-duality of its basic lattice as in Figs. \ref{fig2} and \ref{fig6}.  

\section{Ising and Potts models on random planar lattices}
In this section, we evaluate the critical points for the several spin models on the random planar lattice as in Fig. \ref{fig2}, mixture of the triangular and hexagonal lattices, as a benchmark test of the present method.

\begin{figure}[tbp]
\begin{center}
\includegraphics[width=60mm]{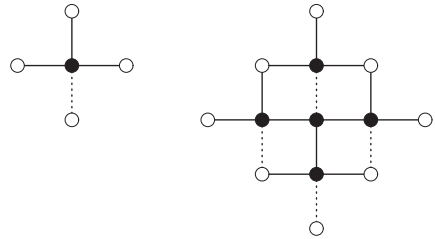}
\end{center}
\caption{The unit cluster for the random planar lattice $s=1$ (left panel).
The right panel denotes the larger cluster $s=2$ to estimate more precise location of the critical points.
The same symbols used as in Fig. \ref{fig1}.}
\label{fig9}
\end{figure}

Let us perform the duality analysis with the real-space renormalization by use of several clusters as in Fig. \ref{fig9}.
By taking the cluster as the basic platform to estimate the location of the critical point, we can deal with the uniform modifications on the random planar lattice. 
We impose that $J_{ij}$ on the remaining bonds without modification is unity in this section, in order to check that our method works well.
In the next section, we apply our method to the spin glass model by introducing another quenched randomness on such a remaining bond $J_{ij}$.

First we take the $q$-state Potts model on the random planar lattice.
We obtain the following formula from Eq. (\ref{MCP}) by use of the four-bond cluster as in Fig. \ref{fig9} (left), namely $s=1$ cluster, to estimate the location of the critical points for the Potts model on the random planar lattice:
\begin{eqnarray}\nonumber
& & 2\log q = p_{\rm mix}\log\left\{\frac{q(1 + (q-1)e^{-\beta_c J})^3}{1+(q-1)e^{-3\beta_c J}}\right\}\\ \nonumber
& & \quad + (1 - p_{\rm mix}) \\ \nonumber
& & \qquad \times \log\left\{(1 + (q-1) e^{-\beta_c J})^3+(q-1)(1-e^{-\beta_c J})^{3}\right\}.\\ \label{Formula1}
\end{eqnarray}
The detailed calculation is shown in Appendix \ref{AP1}.
The special case with $q=2$ corresponds to the critical point for the Ising model through $T_c=2/\beta_c$.
In addition to this case, we list the critical points for the $q=3$ and $q=4$ Potts models by $T_c=1/\beta_c$ in Table \ref{table1}.
The duality relation for the mutually dual pair as confirmed in Figs. \ref{fig3} and \ref{fig4}, $(p_{\rm mix},\exp(-\beta_c J)) \to (1-p_{\rm mix},(\exp(-\beta_c J)-1)/(1+(q-1)\exp(-\beta_c J))$, is satisfied for any value of $p_{\rm mix}$ and $q$.
We can estimate the bond-percolation threshold on the random planar lattice through $p^{\rm th}_c = \exp(-\beta_c J)-1$ on the random planar lattice via a technical limit of $q \to 1$ \cite{Nishimori2011}.
The obtained results are also listed in Table \ref{table1}.
\begin{table}[htbp]
\begin{center}
\begin{tabular}{l|ccc|c}
\hline
$p_{\rm mix}$ & $T_c$ (Ising) & $T_c$ ($q=3$) & $T_c$ ($q=4$) & $p^{\rm th}_c$ \\ \hline
$0.00$ & $3.64096$ & $1.58492$ & $1.44270$ & $0.34730$\\ 
$0.10$ & $3.32027$ & $1.44672$ & $1.31766$ & $0.37422$\\ 
$0.20$ & $3.02041$ & $1.31759$ & $1.20097$ & $0.40334$\\ 
$0.30$ & $2.74404$ & $1.19874$ & $1.09372$ & $0.43436$\\ 
$0.40$ & $2.49324$ & $1.09104$ & $0.99669$ & $0.46680$\\ 
$0.50$ & $2.26919$ & $0.99497$ & $0.91024$ & $0.50000$\\ 
$0.60$ & $2.07194$ & $0.91048$ & $0.83425$ & $0.53320$\\ 
$0.70$ & $1.90046$ & $0.83707$ & $0.76824$ & $0.56564$\\ 
$0.80$ & $1.75281$ & $0.77389$ & $0.71140$ & $0.59666$\\ 
$0.90$ & $1.62647$ & $0.71984$ & $0.66279$ & $0.62578$\\ 
$1.00$ & $1.51865$ & $0.67376$ & $0.62134$ & $0.65270$\\ \hline
\end{tabular}%
\end{center}
\caption{Critical temperatures for the Ising and several ($q=3$ and $4$) Potts models, and the bond-percolation thresholds on the random planar lattice.
The extreme limits $p_{\rm mix} = 0$ and $p_{\rm mix} = 1$ indicate the cases on the triangular and hexagonal lattices, respectively.}
\label{table1}
\end{table}
The well-known exact solutions for the triangular and hexagonal lattices are properly reproduced with $p_{\rm mix}=0$ and $p_{\rm mix}=1$, respectively.
We confirm the Kesten's duality relation, which is the summation of $p^{th}_c$ for $p_{\rm mix}$ and $1-p_{\rm mix}$ should be unity for mutually dual lattices \cite{Kesten1982}.

\section{Random-bond Ising model on random planar lattices}
Next, we apply our duality analysis to the spin glass model with competing interactions on the random planar lattice given by the mixture of the triangular and hexagonal lattices.

Let us perform the duality analysis on the spin glass model on the random planar lattice by introducing the competing interactions on the remaining three bonds on the unit cluster.
We restrict ourselves on the case of the random bond Ising model (o.e. $q=2$) with the following Hamiltonian for simplicity
\begin{equation}
H = - \sum_{\langle ij \rangle}J_{ij}S_iS_j,
\end{equation}
where $S_i = \pm 1$ is the Ising spin hereafter.
The summation is taken over all the bonds on the square lattice.
The interaction $J_{ij}$ remaining after the bond dilution and contraction follows the bimodal distribution as 
\begin{equation}
P(J_{ij}) = p \delta(J_{ij} + J) + (1-p)\delta(J_{ij}-J) = \frac{{\rm e}^{\beta_p J_{ij}}}{2\cosh \beta_p J},\label{dis}
\end{equation}
where $p$ denotes the concentration of the antiferromagnetic interactions.
Here $\beta_p$ is given by $\exp(-2\beta_pJ) = p/(1-p)$.
Notice that we can generalize our scheme to other distribution functions than the above bimodal interactions, such as the multi-component distribution and the Gaussian distribution for $J_{ij}$.
We evaluate the special critical point on the Nishimori line $\beta=\beta_p$, the multicritical point \cite{Nishimori1981,HNBook}.
We express the location of the multicritical point by use of the concentration of the antiferromagnetic interactions $p_c$.
The estimated values $p_c^{(1)}$ and $p_c^{(2)}$ by our duality analysis (i.e. from Eq. (\ref{MCP})) are listed in Table \ref{table2}.
The superscripts denote the size of the cluster used in evaluation of the multicritical points.
The detailed calculation is given in Appendix. \ref{AP2}.

The value of $p_c^{(1)}$ for $p_{\rm mix}=0$ is consistent with the previous result obtained by the combination of the conventional duality and the star-triangle transformation, while the value obtained for $p_{\rm mix}=1$ is slightly different \cite{Nishimori2006}.
The reason for this discrepancy on the hexagonal lattice ($p_{\rm mix}=1$) is that the cluster we used here is slightly different from that in the previous study \cite{Nishimori2006}.
It is known that the results depend on the choice of the cluster in the case of the spin glasses \cite{Ohzeki2009a}.
As discussed in Refs. \cite{Ohzeki2008,Ohzeki2009a,Ohzeki2011a}, if we estimate the critical points by the duality analysis around $T_c$ of the pure Ising model, we can check how well the duality analysis works by comparing to the well-known exact solution for weak randomness .
The values of the critical points around $T_c$ becomes closer to the exact solution by taking larger clusters to obtain the more precise estimations.

Our duality analysis in conjunction with the gauge transformation and the replica method as in Ref. \cite{Ohzeki2009b} can also show absence of the spin glass transition in the symmetric distribution ($p=1/2$) on the random planar lattice.

The random-bond Ising model on the random planer lattice has a close relationship with theoretical limitation of a class of quantum error correction codes called surface codes \cite{Dennis2002,Kitaev2003}.
Two types of the errors, namely bit- and phase-flip errors, on the qubits are associated with the antiferromagnetic interactions on the original and dual lattices, respectively.
The frustrations of the system are related to the error syndromes, which are used to infer the locations of the errors.
As naturally expected, if the error rate, say $p$, of the physical qubits is increased, we cannot store quantum information reliably, which can be viewed that increase of frustration yields the paramagnetic configuration.
This drastic change on the correctability of the error in quantum error correction is rigorously mapped onto the phase transition from the ordered state to disordered one in spin glasses \cite{Dennis2002}.
Thus the multicritical point of the random-bond Ising model corresponds to the theoretical limitation of the surface code, namely optimal error threshold, in the context of quantum error correction.
The detailed description of the relation between the random-bond Ising model and the surface code is given in Appendix \ref{AP3}.

The results in Table \ref{table2} for the random-bond Ising model correspond to the optimal error thresholds of the surface codes defined on the random planer lattices.
They show fairly good agreement with those obtained by a suboptimal method \cite{Loss2012}.
For instance, the suboptimal method gives $p_c = 0.1585$ for $p_{\rm mix}=0$ and $p_c = 0.0645$ for $p_{\rm mix} = 1$, while our duality analysis estimated $p_c^{(2)} = 0.16409$ and $p_c^{(2)}=0.06768$, respectively.
The discrepancies between them are around $4-5 \%$.
Notice that the suboptimal decoding can not achieve the optimal error thresholds which we estimate, since the suboptimal method estimates the critical points in the ground state, $\beta \to \infty$ \cite{HNBook,Dennis2002}.
We can also check the performance of the surface codes on the random planar lattices by computing the following quantity,
\begin{equation}
G(p_c,p_c^*) = H(p_c) + H(p^*_c),
\end{equation}
where $H(p)$ is a binary entropy and $p_c$ and $p_c^*$ indicate the optimal thresholds for $p_{\rm mix}$ and $1-p_{\rm mix}$, respectively, where the argument $p_{\rm mix}$ is omitted for simplicity.
This quantity is related to the quantum Gilbert-Varshamov (q-GV) bound through $R \le 1 - G(p_c)$ in the limit of zero asymptotic rate $R \to 0$ \cite{Dennis2002,Steane1996,Gottesman2003}.
If $G(p_c)$ does not exceed unity, the existence of the good error correcting code is guaranteed.  
The estimate for $p_{\rm mix}=0$ exceeds the q-GV bound as shown in Table \ref{table2}.
However if we use a larger cluster as in Fig. \ref{fig5}, we obtain smaller values for any $p_{\rm mix}$.
We expect that the further analyses by use of the large cluster can give more precise results, which would be consequently below but very close to unity as shown in Refs. \cite{Nishimori2006,Ohzeki2009a}.
\begin{table}[htbp]
\begin{center}
\begin{tabular}{l|cc||cc}
\hline
$p_{\rm mix}$ & $p^{(1)}_c$ & $G(p^{(1)}_c,p^{*(1)}_c)$ &$p^{(2)}_c$& $G(p^{(2)}_c,p^{*(2)}_c)$ \\ \hline
$0.00$ & $0.16419$& $1.00144$ & $0.16409$ & $1.00038$ \\ 
$0.10$ & $0.15323$& $0.99989$ & $0.15308$ & $0.99897$ \\ 
$0.20$ & $0.14207$& $0.99844$ & $0.14191$ & $0.99767$ \\ 
$0.30$ & $0.13091$& $0.99727$ & $0.13075$ & $0.99656$ \\ 
$0.40$ & $0.11994$& $0.99644$ & $0.11982$ & $0.99587$ \\ 
$0.50$ & $0.10940$& $0.99618$ & $0.10930$ & $0.99560$ \\ 
$0.60$ & $0.09945$& --        & $0.09938$ & --        \\ 
$0.70$ & $0.09026$& --        & $0.09018$ & --        \\ 
$0.80$ & $0.08189$& --        & $0.08179$ & --        \\ 
$0.90$ & $0.07437$& --        & $0.07422$ & --        \\ 
$1.00$ & $0.06768$& --        & $0.06746$ & --        \\ \hline
\end{tabular}%
\end{center}
\caption{Optimal error thresholds on the random planar lattice.
The superscript expresses the cluster as in Fig. \ref{fig9} used to estimate the threshold values. }
\label{table2}
\end{table}

\section{Summary}
We perform the duality analysis on the random planar lattices, which consist of various types of the quenched randomness, namely, random modification of the square lattice by bond dilution and contraction, and competing interactions on the resulting random planar lattices.
The duality analysis with the real-space renormalization on the square lattice leads us to the precise locations of the critical points on the random planar lattice.
The known results for the triangular and hexagonal lattices by the star-triangle transformation are properly reproduced in extreme cases of the present framework.
In this sense, the present work is a natural extension of the star-triangle transformation.
By applying our duality analysis, we can also obtain several optimal error thresholds of the surface codes defined on the random planar lattices.

The realm of quantum information has been rapidly spread its range out.
The model, which we deal with, is yielded from the context of quantum information \cite{Loss2012} but the method to analyze it has developed in a different branch of physics, statistical mechanics.
We hope that such fascinating innovations laid across several fields of physics continue to be generated through various studies on quantum information.

\begin{acknowledgements}
We thank the fruitful discussions in the Kinki University through ``Symposium on Quantum Computing, Thermodynamics, and Statistical Physics".
We acknowledge the comments from R. M. Ziff on the correspondence of our result on the bond-percolation thresholds to those from the exact solution as shown in Appendix \ref{AP1}. 
This work was partially supported by MEXT in Japan, Grant-in-Aid for Young
Scientists (B) No.24740263.
\end{acknowledgements}

\appendix 
\section{Derivation of Eq.  (\ref{Formula1}) from Eq. (\ref{MCP})}\label{AP1}
We evaluate Eq. (\ref{MCP}) for the case of the $q$-state Potts model only with the single bond under the bond dilution and contraction on the $s=1$ cluster.
The renormalized Boltzmann factor in the original partition function is written as 
\begin{equation}
x_0^{(1)} = \left[ \left\{ \sum_{S_0} \prod_{i}\left(1+v_i\delta(S_0)\right) \right\}^n\right],
\end{equation}
where $S_0$ is the internal spin summed over, $i$ indicates each bond surrounding the internal spin $S_0$, $n$ is the number of replicas, and $v_i = \exp(\beta J_i)-1$ depending on the quenched disorder.
The product $\prod_i$ runs over four bonds surrounding the internal spin.
On the other hand, the dual renormalized Boltzmann factor is given by 
\begin{equation}
x_0^{*(1)} = \left[  \left\{ \sum_{S_0}\prod_{i}\left(\frac{v_i }{\sqrt{q}}\right)\left(1+\frac{q}{v_i}\delta(S_0)\right) \right\}^n\right].
\end{equation}
After equating both of the quantities as in Eq. (\ref{MCP}), we take $n \to 0$ following the recipe of the replica method.
We reach the generic formula
\begin{equation}
\left[ \log \left( \frac{(q-1)\prod_{i}v_i+\prod_i(q + v_i)}{(q-1)+\prod_i(1 + v_i)} \right)\right] = 2 \log q.
\end{equation}
We set the specific bond $l$ following the distribution
\begin{equation}
P(J_{l}) = p_{\rm mix}\delta(J_{l}) + (1-p_{\rm mix})\delta(J_{l}-\infty),\label{disR}
\end{equation}
while the remaining bonds are constant $J_{i}=J$ for $i \neq l$.
We then obtain 
\begin{eqnarray}\nonumber
&&p_{\rm mix} \log \left( \frac{q (q+v)^3}{(q-1)+(1 + v)^3} \right) \\ \nonumber
&&\quad +(1-p_{\rm mix})\log \left( \frac{(q-1)v^3+(q + v)^3}{(1 + v)^3} \right) \\
&& \qquad = 2 \log q .
\end{eqnarray}
This equality is reduced to Eq. (\ref{Formula1}) by substituting $v=\exp(\beta J)-1$.
If we apply the duality transformation as $v^* \to q/v$, we find the same formula being valid for $p_{\rm mix} \to 1-p_{\rm mix}$.

The limit of $q\to 1$ gives the bond-percolation thresholds on the random planar lattice as in the literature \cite{Nishimori2011}.
The leading order in $\epsilon$ with $q = 1+\epsilon$ gives
\begin{eqnarray}\nonumber
&&p_{\rm mix} \log \left(  1+ \epsilon +\frac{3\epsilon}{1 + v} - \frac{\epsilon}{(1 + v)^3} \right) \\ \nonumber
&&\quad +(1-p_{\rm mix})\log \left( 1+  \frac{3\epsilon}{1 + v} +  \frac{\epsilon v^3}{(1 + v)^3} \right) \\
&& \qquad = 2 \log (1+\epsilon).
\end{eqnarray}
The above relation is reduced to, through $p^{\rm th}_c = v_c/(1+v_c)$,
\begin{equation}
1-3(1-p_{\rm mix})p^{\rm th}_c  - 3p_{\rm mix}(p^{\rm th}_c)^2 + (p^{\rm th}_c)^3=0.
\end{equation}
This equality can be reproduced from the following exact solution on the general formula for the bond-percolation problem \cite{Ziff2006}
\begin{eqnarray}\nonumber
& & C(p,r,s,t) = 1 -pr -ps -rs -pt -rt -st \\
& & \quad +prs +prt+rst+pst = 0, \label{Ceq}
\end{eqnarray}
where $p,r,s$ and $t$ denote the inhomogenous probabilities for the occupied bonds on the unit cell with four bonds of the square lattice.
This equality was originally conjectured \cite{Wu1979} from several evidence and confirmed numerically in high precisions. It was also derived in a different way \cite{Scullard2008}.
Our case corresponds to the case with $p=r=s=p^{\rm th}_c$ and $t=1-p_{\rm mix}$.
This coincidence supports to our theory for the bond-percolation thresholds at least.

\section{Evaluation of $p^{(1)}_{c}$}\label{AP2}
Similarly to the $q$-state Potts model without disorder in interactions, we evaluate the formula to estimate the location of the critical points for the case with disorder in interactions.
The renormalized Boltzmann factor in the original partition function is written as 
\begin{equation}
x_0^{(1)} = \left[ \left\{ \sum_{S_0} \prod_{i}\exp(\beta J_{i}S_0) \right\}^n\right].
\end{equation}
The dual renormalized Boltzmann factor is given by 
\begin{equation}
x_0^{*(1)} = \left[ \left\{ \left(\frac{1 }{4}\right)\sum_{S_0} \prod_{i}\left({\rm e}^{\beta J_{i}}+{\rm e}^{-\beta J_{i}}S_0 \right) \right\}^n\right].
\end{equation}
Taking $n \to 0$ in Eq. (\ref{MCP}), we obtain the following formula
\begin{eqnarray}\nonumber
&&\left[ \log \left( \frac{ \prod_{i}2\cosh \beta J_{i}}{2\cosh \sum_i \beta J_i}  \right) \left(1 + \prod_i\tanh \beta J_i\right)\right] = 2 \log 2.\\
\end{eqnarray}
Numerical evaluation of the above equality gives the estimation of $p_c^{(1)}$ in Table \ref{table2}.
Similarly to the previous case, the bond $l$ follows the distribution function Eq. (\ref{disR}). 
For other remaining bonds, the bimodal distribution function, Eq. (\ref{dis}), is assigned as disorder in the interactions. 
It is straightforward to generalize the above analysis for the larger cluster to obtain more precise results as $p_c^{(2)}$ in Table \ref{table2}.
Then we must take the summation over several internal spins and evaluate the configurational average for all bonds on the cluster.
The complexity of the former computation can be reduced by the special technique as the numerical transfer matrix \cite{Morgenstern1979,Nishimori2011}

\section{Relation between spin glass model and quantum error correction}\label{AP3}
Quantum information is fragile against environmental noise
and results in the loss of coherence, that is, {\it decoherence}.
It is one of the main issue for the realization of quantum information processing
to counteract the errors due to decoherence.
In comparison with classical information,
quantum bit, namely {\it qubit}, is subject to not only bit but also phase flip errors represented by Pauli $X$ and $Z$ operators, respectively.
Quantum error correction \cite{Shor1995} is one of the most successful scheme to handle these errors,
which employs multiple physical qubits to encode logical information
into its subspace, so-called {\it code space}.

The code space is defined by the stabilizer operators $\{ S_i \}$,
which are the elements of an Abelian subgroup of the $n$-qubit  Pauli group $\{ -iI,X,Y,Z \} ^{\otimes n}$ \cite{Gottesman1997}.
That is, for any quantum state $|\psi _{L} \rangle $ in the code space,
$S_i | \psi _L \rangle = | \psi _L\rangle $ is satisfied for all $i$.
If an error occurs on the physical qubits, the state is mapped 
onto an orthogonal subspace of the code space.
This means that we can 
detect the occurrence of the error by measuring the eigenvalue $s _i$ of the stabilizer operator $S_i$,
which we call {\it error syndrome}.
Specifically, $s _i =-1$ implies an existence of the error that anticommutes with $S_i$.

\begin{figure}
\centering
\includegraphics[width=85mm]{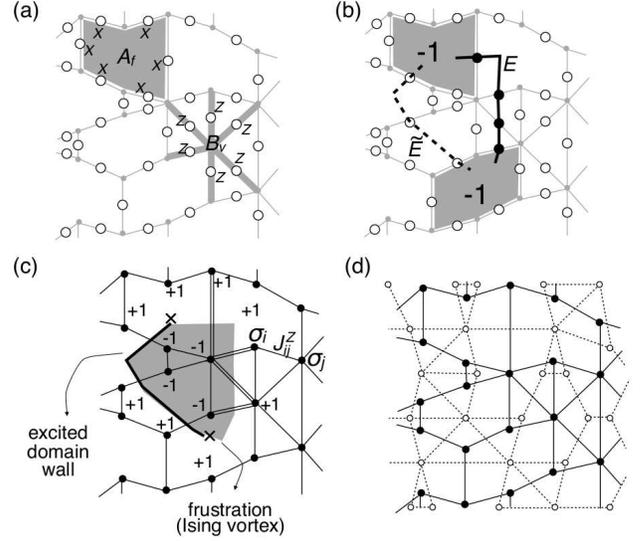}
\caption{(a) A qubit, denoted by the open circle, is located on each edge of the lattice.
The stabilizer operators defined on each face and vertex. (b) $Z$ error chains $E$
and $\tilde E$, which have the same error syndrome. The incorrect error syndromes
are displayed by the gray faces. (c) The associated RBIM with the $Z$ error correction problem. 
An Ising spin is located on each vertex of the lattice.
A coupling constant  $J_{ij}^{Z}=f_{ij} (E)$ is assigned on each edge.
The actual and hypothetical error chains $E$ and $\tilde E$ correspond to the anti-ferromagnetic interactions shown by the double lines. The excited domain wall is shown by the bold line.
The gray-shaded area is a domain of $\sigma _i =-1$. 
The crosses indicate frustrations. 
Equivalently they are also the end points of the excited domain wall, so-called Ising vortexes.
(d) Primal (solid lines and closed circles) and dual (dotted lines and open circles) lattices.
According to the definition of the stabilizer operators shown in (a), the $Z$ and $X$ error correction problems are mapped to the RBIMs on the primal and dual lattices, whose concentrations of anti-ferromagnetic interactions are given by $p_Z$ and $p_X$, respectively. }
\label{ApFig}
\end{figure}

The {\it surface codes} \cite{Dennis2002}, which have a close relationship with the spin glass models,
are a class of quantum error correction codes,
whose stabilizer operators are defined on the surfaces of lattices as follows.
Suppose that a qubit is associated with each edge of a lattice (see Fig. \ref{ApFig} (a)). 
Then, the stabilizer operators are defined 
for each face $f$ and vertex $v$ of the lattice as 
\begin{eqnarray*}
A_f=\prod _{l \in f}X_l, \;\;\;
B_v=\prod _{l \in v} Z_l,
\end{eqnarray*}
where $C_l$ ($C=X,Z$) indicates
the Pauli operators of the qubit located on the edge $l$,
$\prod _{l \in f}$ and $\prod _{l \in v}$ are performed for the qubits (edges) located around the face $f$
and adjacent to the vertex $v$.
The error syndromes $a_f$ and $b_v$ detect $Z$ (phase) and $X$ (bit) errors, respectively.
Note that if a $Z$ error chain, say $E$, occurs on the code space, those error syndromes that are located on the boundary $\partial E$ of the error chain $E$ return the eigenvalue $a_f=-1$ as shown in Fig, \ref{ApFig} (b).
This is also the case for $X$ errors and the error syndrome $b_v$.
Although we, for simplicity, consider only $Z$ errors below, the same argument works for $X$ errors.

In the error correction, we infer the location of the error from the error syndrome $\partial E$.
Suppose that $Z$ errors occur with probability $p_Z$ independently for each qubit.
Conditioned on the error syndrome $\partial E$, the probability of a hypothetical error chain $\tilde E$ that has the same error syndrome is given by
\begin{eqnarray*}
p( \tilde E | \partial E) 
= \left.\left(\frac{p'_Z}{1-p'_Z} \right)^{|B|/2} \prod _{l \in B} \left(\sqrt{\frac{p'_Z}{1-p'_Z}} \right)^{-f_{l}({\tilde E})} \right\rvert_{\partial E = \partial \tilde E_Z}, 
\end{eqnarray*}
where $B$ is the set of edges (bond) of the lattice, $p'_Z$ is the hypothetical probability of the occurrence of $Z$ error for each qubit, and 
\begin{eqnarray*}
f_l (\tilde E)  =
\left\{
\begin{array}{c}
-1 \textrm{ for } l \in E
\\
+1 \textrm{ for } l \notin E
\end{array}
\right.,
\end{eqnarray*}
which indicates the location of the error chain $E$.
(Note that the actual and hypothetical error probabilities $p_Z$ and $p'_Z$ 
are not necessary the same.)
The error correction is done by operating the estimated error chain $\tilde E$ on the state, and hence the net effect on the code space is $\tilde E+ E$. 
If $\tilde E + E$ belongs to the trivial homology class as shown in Fig. \ref{ApFig} (b), the error correction succeeds
(otherwise, $\tilde E + E$ is a non-trivial cycle and hence acts on the code space non-trivially, which results in an logical error \cite{Dennis2002}).
By taking a summation over such $\tilde E$, we obtain the success probability of the error correction
\begin{eqnarray*}
p_{\rm suc} \propto \sum _{\tilde E| \partial E = \partial \tilde  E }  
\exp\left [\beta _Z{ \sum _{l} f_{l}(\tilde E)} \right],
\end{eqnarray*}
where $\beta _Z$ is defined such that $e^{ -\beta _Z } = \sqrt{p'_Z/(1-p'_Z)}$.
In order to simplify the summation over the same homology class, we introduce an Ising spin  $\sigma _{i} \in \{ +1, -1 \}$ on each site $i$ of the lattice.
If $\tilde E+E$ belongs to the trivial homology class, there exists a configuration $\{ \sigma _i \}$
such that $f_l(\tilde E)$ can be expressed as $f_l(\tilde E  )= f_l(E) \sigma _i \sigma _j $ with the sites $i$ and $j$ connected by the edge $l$.
By using this fact, the success probability can be reformulated as 
\begin{eqnarray*}
p_{\rm suc} &\propto& 
\sum _{\{ \sigma _i \}  }  \exp \left[ {  \beta _Z \sum _{\langle ij \rangle }J^{Z}_{ij} \sigma _i \sigma _j} \right],
\end{eqnarray*}
where the edge $l$ are replaced by the pair of sites $ij$, and $f_{ij}(E)$ is simply denoted by $J_{ i j } ^{Z}$.

Now that the relation between the success probability and spin glass model becomes apparent;
the success probability of the $Z$ error correction is nothing but the partition function of the random-bond Ising model (RBIM), whose Hamiltonian is given by $H= - \sum _{\langle ij \rangle} J_{ij}^Z \sigma _i \sigma _j$.
The location of $Z$ errors represented by $J^{Z}_{ i j} = -1$ corresponds to the anti-ferromagnetic interaction due to disorder, whose probability distribution is given by
\begin{eqnarray}
P( J^{Z}_{ i  j} ) =(1-p_Z) \delta ( J^{Z}_{ i  j} -1 )+ p_Z \delta ( J^{Z}_{ i  j} +1 ).
\end{eqnarray}
(Recall that the hypothetical error probability $p'_Z$ is given independently of the actual error probability $p_Z$.)
The error syndrome $a_f=-1$ corresponds to the frustration of Ising interactions. 
Equivalently it is also the end point (Ising vortex) of the excited domain wall as shown in Fig. \ref{ApFig} (c). 

In order to storage quantum information reliably, the success probability has to be exponentially small
by increasing the system size.
While this is achieved with the error probability $p_Z$ smaller than a certain value, so-called {\it accuracy threshold} $p^{\rm th}_Z$, if the error probability is higher than it, the success probability converges to a constant value.
This drastic change of the function $p_{\rm suc}$ comes from a phase transition of the RBIM;
in the ferromagnetic phase quantum error correction succeeds.
Of course, we should take $p_Z= 1/(e^{2\beta _Z} +1) = p'_Z$, that is, the actual and hypothetical error probabilities are the same, in order to perform a better error correction.
This condition corresponds to the Nishimori line \cite{Nishimori1981,HNBook} on the $(\beta _Z,p_Z)$ phase diagram.

The same argument leads that the $X$ error correction is mapped to the RBIM on the dual lattice (see Fig. \ref{ApFig} (d)).
The inverse temperature and concentration of the anti-ferromagnetic interactions are taken as $\beta _X$ and $p_X$ respectively.
For each surface code, we have two RBIMs defined on the primal and dual lattices, whose inverse temperatures and concentrations are $\beta _{Z,X}$ and $p_{Z,X}$ respectively.
Thus, if we modify a surface code on a primal lattice, both RBIMs on the primal and dual lattices are simultaneously changed.
Specifically, the random modification of the stabilizer operators by a systematic removal of the qubits introduced in Ref. \cite{Loss2012} corresponds to the bond dilution and contraction, which have mutually dual actions on the primal and dual lattices.

\bibliography{paper-2_ver2}

\begin{thebibliography}{29}
\expandafter\ifx\csname natexlab\endcsname\relax\def\natexlab#1{#1}\fi
\expandafter\ifx\csname bibnamefont\endcsname\relax
  \def\bibnamefont#1{#1}\fi
\expandafter\ifx\csname bibfnamefont\endcsname\relax
  \def\bibfnamefont#1{#1}\fi
\expandafter\ifx\csname citenamefont\endcsname\relax
  \def\citenamefont#1{#1}\fi
\expandafter\ifx\csname url\endcsname\relax
  \def\url#1{\texttt{#1}}\fi
\expandafter\ifx\csname urlprefix\endcsname\relax\def\urlprefix{URL }\fi
\providecommand{\bibinfo}[2]{#2}
\providecommand{\eprint}[2][]{\url{#2}}

\bibitem[{\citenamefont{Kramers and Wannier}(1941)}]{Kramers1941}
\bibinfo{author}{\bibfnamefont{H.~A.} \bibnamefont{Kramers}} \bibnamefont{and}
  \bibinfo{author}{\bibfnamefont{G.~H.} \bibnamefont{Wannier}},
  \bibinfo{journal}{Phys. Rev.} \textbf{\bibinfo{volume}{60}},
  \bibinfo{pages}{252} (\bibinfo{year}{1941}),
  \urlprefix\url{http://link.aps.org/doi/10.1103/PhysRev.60.252}.

\bibitem[{\citenamefont{Wu and Wang}(1976)}]{Wu1976}
\bibinfo{author}{\bibfnamefont{F.~Y.} \bibnamefont{Wu}} \bibnamefont{and}
  \bibinfo{author}{\bibfnamefont{Y.~K.} \bibnamefont{Wang}},
  \bibinfo{journal}{Journal of Mathematical Physics}
  \textbf{\bibinfo{volume}{17}}, \bibinfo{pages}{439} (\bibinfo{year}{1976}),
  \urlprefix\url{http://link.aip.org/link/?JMP/17/439/1}.

\bibitem[{\citenamefont{Nishimori and Nemoto}(2002)}]{Nishimori2002}
\bibinfo{author}{\bibfnamefont{H.}~\bibnamefont{Nishimori}} \bibnamefont{and}
  \bibinfo{author}{\bibfnamefont{K.}~\bibnamefont{Nemoto}},
  \bibinfo{journal}{Journal of the Physical Society of Japan}
  \textbf{\bibinfo{volume}{71}}, \bibinfo{pages}{1198} (\bibinfo{year}{2002}),
  \urlprefix\url{http://jpsj.ipap.jp/link?JPSJ/71/1198/}.

\bibitem[{\citenamefont{Maillard et~al.}(2003)\citenamefont{Maillard, Nemoto,
  and Nishimori}}]{Mailard2003}
\bibinfo{author}{\bibfnamefont{J.-M.} \bibnamefont{Maillard}},
  \bibinfo{author}{\bibfnamefont{K.}~\bibnamefont{Nemoto}}, \bibnamefont{and}
  \bibinfo{author}{\bibfnamefont{H.}~\bibnamefont{Nishimori}},
  \bibinfo{journal}{Journal of Physics A: Mathematical and General}
  \textbf{\bibinfo{volume}{36}}, \bibinfo{pages}{9799} (\bibinfo{year}{2003}),
  \urlprefix\url{http://stacks.iop.org/0305-4470/36/i=38/a=301}.

\bibitem[{\citenamefont{Takeda et~al.}(2005)\citenamefont{Takeda, Sasamoto, and
  Nishimori}}]{Takeda2005}
\bibinfo{author}{\bibfnamefont{K.}~\bibnamefont{Takeda}},
  \bibinfo{author}{\bibfnamefont{T.}~\bibnamefont{Sasamoto}}, \bibnamefont{and}
  \bibinfo{author}{\bibfnamefont{H.}~\bibnamefont{Nishimori}},
  \bibinfo{journal}{Journal of Physics A: Mathematical and General}
  \textbf{\bibinfo{volume}{38}}, \bibinfo{pages}{3751} (\bibinfo{year}{2005}),
  \urlprefix\url{http://stacks.iop.org/0305-4470/38/i=17/a=004}.

\bibitem[{\citenamefont{Nishimori and Ohzeki}(2006)}]{Nishimori2006}
\bibinfo{author}{\bibfnamefont{H.}~\bibnamefont{Nishimori}} \bibnamefont{and}
  \bibinfo{author}{\bibfnamefont{M.}~\bibnamefont{Ohzeki}},
  \bibinfo{journal}{Journal of the Physical Society of Japan}
  \textbf{\bibinfo{volume}{75}}, \bibinfo{pages}{034004}
  (\bibinfo{year}{2006}),
  \urlprefix\url{http://dx.doi.org/10.1143/JPSJ.75.034004}.

\bibitem[{\citenamefont{Ohzeki et~al.}(2008)\citenamefont{Ohzeki, Nishimori,
  and Berker}}]{Ohzeki2008}
\bibinfo{author}{\bibfnamefont{M.}~\bibnamefont{Ohzeki}},
  \bibinfo{author}{\bibfnamefont{H.}~\bibnamefont{Nishimori}},
  \bibnamefont{and} \bibinfo{author}{\bibfnamefont{A.~N.}
  \bibnamefont{Berker}}, \bibinfo{journal}{Phys. Rev. E}
  \textbf{\bibinfo{volume}{77}}, \bibinfo{pages}{061116}
  (\bibinfo{year}{2008}),
  \urlprefix\url{http://link.aps.org/doi/10.1103/PhysRevE.77.061116}.

\bibitem[{\citenamefont{Ohzeki}(2009{\natexlab{a}})}]{Ohzeki2009a}
\bibinfo{author}{\bibfnamefont{M.}~\bibnamefont{Ohzeki}},
  \bibinfo{journal}{Phys. Rev. E} \textbf{\bibinfo{volume}{79}},
  \bibinfo{pages}{021129} (\bibinfo{year}{2009}{\natexlab{a}}),
  \urlprefix\url{http://link.aps.org/doi/10.1103/PhysRevE.79.021129}.

\bibitem[{\citenamefont{Dennis et~al.}(2002)\citenamefont{Dennis, Kitaev,
  Landahl, and Preskill}}]{Dennis2002}
\bibinfo{author}{\bibfnamefont{E.}~\bibnamefont{Dennis}},
  \bibinfo{author}{\bibfnamefont{A.}~\bibnamefont{Kitaev}},
  \bibinfo{author}{\bibfnamefont{A.}~\bibnamefont{Landahl}}, \bibnamefont{and}
  \bibinfo{author}{\bibfnamefont{J.}~\bibnamefont{Preskill}},
  \bibinfo{journal}{Journal of Mathematical Physics}
  \textbf{\bibinfo{volume}{43}}, \bibinfo{pages}{4452} (\bibinfo{year}{2002}),
  \urlprefix\url{http://link.aip.org/link/?JMP/43/4452/1}.

\bibitem[{\citenamefont{Ohzeki}(2009{\natexlab{b}})}]{Ohzeki2009c}
\bibinfo{author}{\bibfnamefont{M.}~\bibnamefont{Ohzeki}},
  \bibinfo{journal}{Physical Review E} \textbf{\bibinfo{volume}{80}},
  \bibinfo{pages}{011141} (\bibinfo{year}{2009}{\natexlab{b}}),
  \urlprefix\url{http://dx.doi.org/10.1103/PhysRevE.80.011141}.

\bibitem[{\citenamefont{Bombin et~al.}(2012)\citenamefont{Bombin, Andrist,
  Ohzeki, Katzgraber, and Martin-Delgado}}]{Hector2012}
\bibinfo{author}{\bibfnamefont{H.}~\bibnamefont{Bombin}},
  \bibinfo{author}{\bibfnamefont{R.~S.} \bibnamefont{Andrist}},
  \bibinfo{author}{\bibfnamefont{M.}~\bibnamefont{Ohzeki}},
  \bibinfo{author}{\bibfnamefont{H.~G.} \bibnamefont{Katzgraber}},
  \bibnamefont{and} \bibinfo{author}{\bibfnamefont{M.~A.}
  \bibnamefont{Martin-Delgado}}, \bibinfo{journal}{Phys. Rev. X}
  \textbf{\bibinfo{volume}{2}}, \bibinfo{pages}{021004} (\bibinfo{year}{2012}),
  \urlprefix\url{http://link.aps.org/doi/10.1103/PhysRevX.2.021004}.

\bibitem[{\citenamefont{R\"othlisberger
  et~al.}(2012)\citenamefont{R\"othlisberger, Wootton, Heath, Pachos, and
  Loss}}]{Loss2012}
\bibinfo{author}{\bibfnamefont{B.}~\bibnamefont{R\"othlisberger}},
  \bibinfo{author}{\bibfnamefont{J.~R.} \bibnamefont{Wootton}},
  \bibinfo{author}{\bibfnamefont{R.~M.} \bibnamefont{Heath}},
  \bibinfo{author}{\bibfnamefont{J.~K.} \bibnamefont{Pachos}},
  \bibnamefont{and} \bibinfo{author}{\bibfnamefont{D.}~\bibnamefont{Loss}},
  \bibinfo{journal}{Phys. Rev. A} \textbf{\bibinfo{volume}{85}},
  \bibinfo{pages}{022313} (\bibinfo{year}{2012}),
  \urlprefix\url{http://link.aps.org/doi/10.1103/PhysRevA.85.022313}.

\bibitem[{\citenamefont{Domb and Green.}(1972)}]{Domb1972}
\bibinfo{editor}{\bibfnamefont{C.}~\bibnamefont{Domb}} \bibnamefont{and}
  \bibinfo{editor}{\bibfnamefont{M.}~\bibnamefont{Green.}}, eds.,
  \emph{\bibinfo{title}{Phase Transitions and Critical Phenomena. Vol.1: Exact
  results.}} (\bibinfo{publisher}{Academic Press London.},
  \bibinfo{year}{1972}), ISBN \bibinfo{isbn}{0122203011}.

\bibitem[{\citenamefont{Wu}(1982)}]{Wu1982}
\bibinfo{author}{\bibfnamefont{F.~Y.} \bibnamefont{Wu}}, \bibinfo{journal}{Rev.
  Mod. Phys.} \textbf{\bibinfo{volume}{54}}, \bibinfo{pages}{235}
  (\bibinfo{year}{1982}),
  \urlprefix\url{http://link.aps.org/doi/10.1103/RevModPhys.54.235}.

\bibitem[{\citenamefont{Nishimori and Ortiz}(2011)}]{Nishimori2011}
\bibinfo{author}{\bibfnamefont{H.}~\bibnamefont{Nishimori}} \bibnamefont{and}
  \bibinfo{author}{\bibfnamefont{G.}~\bibnamefont{Ortiz}},
  \emph{\bibinfo{title}{Elements of Phase Transitions and Critical Phenomena
  (Oxford Graduate Texts)}} (\bibinfo{publisher}{Oxford University Press, USA},
  \bibinfo{year}{2011}), ISBN \bibinfo{isbn}{0199577226}.

\bibitem[{\citenamefont{Ohzeki et~al.}(2011)\citenamefont{Ohzeki, Thomas,
  Katzgraber, Bombin, and Martin-Delgado}}]{Ohzeki2011a}
\bibinfo{author}{\bibfnamefont{M.}~\bibnamefont{Ohzeki}},
  \bibinfo{author}{\bibfnamefont{C.~K.} \bibnamefont{Thomas}},
  \bibinfo{author}{\bibfnamefont{H.~G.} \bibnamefont{Katzgraber}},
  \bibinfo{author}{\bibfnamefont{H.}~\bibnamefont{Bombin}}, \bibnamefont{and}
  \bibinfo{author}{\bibfnamefont{M.~A.} \bibnamefont{Martin-Delgado}},
  \bibinfo{journal}{Journal of Statistical Mechanics: Theory and Experiment}
  \textbf{\bibinfo{volume}{2011}}, \bibinfo{pages}{P02004}
  (\bibinfo{year}{2011}),
  \urlprefix\url{http://stacks.iop.org/1742-5468/2011/i=02/a=P02004}.

\bibitem[{\citenamefont{Kesten}(1982)}]{Kesten1982}
\bibinfo{author}{\bibfnamefont{H.}~\bibnamefont{Kesten}},
  \emph{\bibinfo{title}{Percolation theory for mathematicians}}, Progress in
  probability and statistics (\bibinfo{publisher}{Birkh{\"a}user},
  \bibinfo{year}{1982}), ISBN \bibinfo{isbn}{9783764331078}.

\bibitem[{\citenamefont{Nishimori}(1981)}]{Nishimori1981}
\bibinfo{author}{\bibfnamefont{H.}~\bibnamefont{Nishimori}},
  \bibinfo{journal}{Progress of Theoretical Physics}
  \textbf{\bibinfo{volume}{66}}, \bibinfo{pages}{1169} (\bibinfo{year}{1981}),
  \urlprefix\url{http://ptp.ipap.jp/link?PTP/66/1169/}.

\bibitem[{\citenamefont{Nishimori}(2001)}]{HNBook}
\bibinfo{author}{\bibfnamefont{H.}~\bibnamefont{Nishimori}},
  \emph{\bibinfo{title}{Statistical physics of spin glasses and information
  processing : an introduction}} (\bibinfo{publisher}{Oxford University Press},
  \bibinfo{address}{Oxford New York}, \bibinfo{year}{2001}), ISBN
  \bibinfo{isbn}{0198509413}.

\bibitem[{\citenamefont{Ohzeki and Nishimori}(2009)}]{Ohzeki2009b}
\bibinfo{author}{\bibfnamefont{M.}~\bibnamefont{Ohzeki}} \bibnamefont{and}
  \bibinfo{author}{\bibfnamefont{H.}~\bibnamefont{Nishimori}},
  \bibinfo{journal}{Journal of Physics A: Mathematical and Theoretical}
  \textbf{\bibinfo{volume}{42}}, \bibinfo{pages}{332001}
  (\bibinfo{year}{2009}),
  \urlprefix\url{http://stacks.iop.org/1751-8121/42/i=33/a=332001}.

\bibitem[{\citenamefont{Kitaev}(2003)}]{Kitaev2003}
\bibinfo{author}{\bibfnamefont{A.}~\bibnamefont{Kitaev}},
  \bibinfo{journal}{Annals of Physics} \textbf{\bibinfo{volume}{303}},
  \bibinfo{pages}{2 } (\bibinfo{year}{2003}), ISSN \bibinfo{issn}{0003-4916},
  \urlprefix\url{http://www.sciencedirect.com/science/article/pii/S00034916020%
00180}.

\bibitem[{\citenamefont{Steane}(1996)}]{Steane1996}
\bibinfo{author}{\bibfnamefont{A.~M.} \bibnamefont{Steane}},
  \bibinfo{journal}{Phys. Rev. Lett.} \textbf{\bibinfo{volume}{77}},
  \bibinfo{pages}{793} (\bibinfo{year}{1996}),
  \urlprefix\url{http://link.aps.org/doi/10.1103/PhysRevLett.77.793}.

\bibitem[{\citenamefont{Gottesman and Lo}(2003)}]{Gottesman2003}
\bibinfo{author}{\bibfnamefont{D.}~\bibnamefont{Gottesman}} \bibnamefont{and}
  \bibinfo{author}{\bibfnamefont{H.-K.} \bibnamefont{Lo}},
  \bibinfo{journal}{Information Theory, IEEE Transactions on}
  \textbf{\bibinfo{volume}{49}}, \bibinfo{pages}{457 } (\bibinfo{year}{2003}),
  ISSN \bibinfo{issn}{0018-9448}.

\bibitem[{\citenamefont{Ziff}(2006)}]{Ziff2006}
\bibinfo{author}{\bibfnamefont{R.~M.} \bibnamefont{Ziff}},
  \bibinfo{journal}{Phys. Rev. E} \textbf{\bibinfo{volume}{73}},
  \bibinfo{pages}{016134} (\bibinfo{year}{2006}),
  \urlprefix\url{http://link.aps.org/doi/10.1103/PhysRevE.73.016134}.

\bibitem[{\citenamefont{Wu}(1979)}]{Wu1979}
\bibinfo{author}{\bibfnamefont{F.~Y.} \bibnamefont{Wu}},
  \bibinfo{journal}{Journal of Physics C: Solid State Physics}
  \textbf{\bibinfo{volume}{12}}, \bibinfo{pages}{L645} (\bibinfo{year}{1979}),
  \urlprefix\url{http://stacks.iop.org/0022-3719/12/i=17/a=002}.

\bibitem[{\citenamefont{Scullard and Ziff}(2008)}]{Scullard2008}
\bibinfo{author}{\bibfnamefont{C.~R.} \bibnamefont{Scullard}} \bibnamefont{and}
  \bibinfo{author}{\bibfnamefont{R.~M.} \bibnamefont{Ziff}},
  \bibinfo{journal}{Phys. Rev. Lett.} \textbf{\bibinfo{volume}{100}},
  \bibinfo{pages}{185701} (\bibinfo{year}{2008}),
  \urlprefix\url{http://link.aps.org/doi/10.1103/PhysRevLett.100.185701}.

\bibitem[{\citenamefont{Morgenstern and Binder}(1979)}]{Morgenstern1979}
\bibinfo{author}{\bibfnamefont{I.}~\bibnamefont{Morgenstern}} \bibnamefont{and}
  \bibinfo{author}{\bibfnamefont{K.}~\bibnamefont{Binder}},
  \bibinfo{journal}{Phys. Rev. Lett.} \textbf{\bibinfo{volume}{43}},
  \bibinfo{pages}{1615} (\bibinfo{year}{1979}),
  \urlprefix\url{http://link.aps.org/doi/10.1103/PhysRevLett.43.1615}.

\bibitem[{\citenamefont{Shor}(1995)}]{Shor1995}
\bibinfo{author}{\bibfnamefont{P.~W.} \bibnamefont{Shor}},
  \bibinfo{journal}{Phys. Rev. A} \textbf{\bibinfo{volume}{52}},
  \bibinfo{pages}{R2493} (\bibinfo{year}{1995}),
  \urlprefix\url{http://link.aps.org/doi/10.1103/PhysRevA.52.R2493}.

\bibitem[{\citenamefont{Gottesman}(1997)}]{Gottesman1997}
\bibinfo{author}{\bibfnamefont{D.}~\bibnamefont{Gottesman}},
  \bibinfo{journal}{Ph.D. thesis, California Institute of Technology,
  arXiv:quant-ph/9705052}  (\bibinfo{year}{1997}).

\end{thebibliography}


\end{document}